\documentclass[nohyper,10pt,letterpaper]{JHEP3}

\usepackage[centertags]{amsmath}
\usepackage{amssymb}
\usepackage{cite}
\usepackage{graphicx}
\usepackage{eepic}
\usepackage{epic} 
\usepackage[dvips]{epsfig}

\title{The Incompressible Non-Relativistic Navier-Stokes Equation from Gravity}

\author{Sayantani Bhattacharyya$^a$\footnote{sayanta@theory.tifr.res.in}, 
Shiraz Minwalla $^a$\footnote{minwalla@theory.tifr.res.in} and 
Spenta R. Wadia$^a$\footnote{wadia@theory.tifr.res.in} \\
\small{\emph{$^{a}$Department of Theoretical Physics,Tata Institute of Fundamental Research,}} \\
\small{\emph{Homi Bhabha Rd, Mumbai 400005, India}} \\ 
}

\abstract{We note that the equations of relativistic hydrodynamics reduce 
to the incompressible Navier-Stokes equations in a particular scaling limit. 
In this limit boundary metric fluctuations of the underlying relativistic 
system turn into a forcing function identical to 
the action of a background electromagnetic field on the effectively 
charged fluid. We demonstrate that special conformal symmetries of the 
parent relativistic theory descend to `accelerated boost' symmetries of 
the Navier-Stokes equations, uncovering a conformal symmetry
structure of these equations. Applying our scaling limit  to holographically 
induced fluid dynamics, we find gravity dual descriptions of an arbitrary
solution of the forced non-relativistic incompressible Navier-Stokes 
equations. In the holographic context we also find a simple forced steady 
state shear solution to the Navier-Stokes equations, and demonstrate 
that this solution turns unstable at high enough Reynolds numbers, 
indicating a possible eventual transition to turbulence.}

\keywords{}
\preprint{TIFR/TH/08-40}

\begin{document}

\section{Introduction and Discussion}

The flow of non-relativistic fluids is described by the Navier-Stokes equations
\begin{equation}\label{ns} \begin{split}
{\dot {\vec v} }+ {\vec v}. \nabla  {\vec v}& = -{\vec \nabla} P 
+ \nu \nabla^2 {\vec v} + {\vec f} \\
{\vec \nabla}. {\vec v}&=0\\
\end{split}
\end{equation}
where ${\vec v}$ is the fluid velocity, $P$ the fluid pressure, $\nu$ the 
shear viscosity and ${\vec f}$ an externally specified forcing function.
Although these equations describe a wide variety of natural phenomena 
(see e.g. \cite{1959flme.book.....L} ) and have been intensively studied
for almost two centuries, their extremely rich phenomenology remains very
poorly understood. In particular, most fluid flows go turbulent at high
Reynolds number, i.e. in the regime in which the viscous fluid term is 
negligible compared to the nonlinear convective term in \eqref{ns}. Although
turbulent flows appear complicated and statistical in nature, it has been 
suggested (see e.g. \cite{Polyakov:1992yw} ) that these flows are in fact
governed by a new and simple universal mathematical structure analogous to a
fixed point of the renormalization group flow equations. A completely new
angle on fluid dynamics could well be needed in order to uncover such a structure.

Recent investigations\cite{Son:2007vk, Bhattacharyya:2007jc,
Loganayagam:2008is,VanRaamsdonk:2008fp, Dutta:2008gf, Bhattacharyya:2008xc, 
Bhattacharyya:2008ji, Haack:2008cp,Erdmenger:2008rm,
Banerjee:2008th, Bhattacharyya:2008mz}, within the framework of the AdS/CFT
correspondence of string theory\cite{Maldacena:1997re} have revealed an
initially surprising relationship between the vacuum  equations of Einstein
gravity in an asymptotically locally AdS$_{d+1}$ space and the equations
of hydrodynamics in $d$ dimensions. More concretely these papers study a
class of regular, long wavelength  locally asymptotically AdS$_{d+1}$ solutions
to the vacuum Einstein equations with a negative cosmological constant. These
solutions are shown to be in one to one correspondence with solutions of the 
$d$ dimensional hydrodynamical equations $\nabla_\mu T^{\mu\nu}=0$. In the 
last equation the stress tensor $T^{\mu\nu}$ is a holographically determined
functional of a $d$ dimensional fluid velocity $u^\mu$ and temperature $T$.
In the long distance limit under consideration it is appropriate to expand
the stress tensor in a power series in the boundary derivatives of the velocity
and temperature fields. Schematically
\begin{equation}\label{expst}
 T^{\mu \nu} = \sum_{n=0}^\infty T^{d-n} T_n^{\mu\nu}
\end{equation}
where $T$ is the local fluid temperature and $T_n^{\mu\nu}$ is a local 
function of the fluid velocity and temperature of $n^{th}$ order in spacetime 
derivatives. The expressions for $T^{\mu\nu}_n$ for $n \leq 2$ have been 
explicitly determined in the references cited above (see 
\cite{Bhattacharyya:2008mz} for the most general result) and constitute
a relativistic generalization of the incompressible Navier-Stokes stress 
tensor. In summary, classical asymptotically AdS$_{d+1}$ gravity is `dual' to 
relativistic generalizations of the Navier-Stokes equations at long distance 
and time scales. 

Many theoretical and experimental investigations of fluid dynamics 
study the actual incompressible Navier-Stokes equations \eqref{ns}. 
It is consequently of interest to find a dual description of the 
Navier-Stokes equations \eqref{ns} themselves rather than their their 
relativistic generalizations. This may be simply achieved by taking 
the appropriate limit of the results of 
\cite{Bhattacharyya:2007jc,Loganayagam:2008is,VanRaamsdonk:2008fp, 
Dutta:2008gf, Bhattacharyya:2008xc, Bhattacharyya:2008ji, Haack:2008cp,
Erdmenger:2008rm, Banerjee:2008th, Bhattacharyya:2008mz}, and this limit 
is the topic of our note.

In order to find the gravitational description dual to \eqref{ns}, 
we adopt a two step procedure. First, purely at the level of fluid dynamics
we make a straightforward and possibly well known observation. We note that the non-relativistic incompressible Navier-Stokes equations \eqref{ns} are the precise and universal outcome of a particular combined scaling limit (one in which
we scale to long distances, long times, low speeds and low amplitudes in
a coordinated fashion) applied to any reasonable relativistic equations
of hydrodynamics, i.e. the hydrodynamics a relativistic fluid with any
reasonable equation of state\footnote{For instance the fluid that is dual to
gravity studied below whose equation of state is dictated by conformal 
invariance.}.The equations of fluid dynamics become non-relativistic at
low speeds for the usual reason, and also become effectively incompressible,
as we go to velocities much lower than the speed of sound (see for instance \cite{1959flme.book.....L}).

To be more specific we show that the equations of hydrodynamics reduce, in
a precise fashion, to the incompressible non-relativistic Navier-Stokes
equations \eqref{ns} under the limit 
\begin{equation}\label{scaling}\begin{split}
\delta x& \sim \frac{1}{T \epsilon}\\
\delta t&\sim  \frac{1}{T \epsilon^2}\\
v^i & \sim   \epsilon\\
\delta P & \sim  T^d \epsilon^2\\
\epsilon & \to 0
\end{split}
\end{equation}
where $\delta x$ is the spatial length scale, $\delta t$ the temporal 
scale while  $v^i$ and $\delta P$  represent estimates of the magnitude 
of velocity and pressure fluctuations about an ambient configuration of 
equilibrium fluid at rest. The rough contours of this choice of scaling are 
quite intuitive. It is clear we have to scale to long distances to be 
in the fluid dynamical regime.  The dispersion relation for shear waves, 
$\omega = i \nu k^2$ suggests that time intervals should scale like 
spatial intervals squared\footnote{The scaling to arbitrarily low 
velocities projects out sound waves with dispersion relation $\omega \propto k
$.}. Scalings of distances and time intervals determine the scaling law
for velocities. Finally the pressure variations are scaled appropriately to
ensure that they cannot accelerate the fluid into velocities outside 
this scaling limit. 

As we have explained, the scaling  \eqref{scaling} of the equations of 
relativistic fluid mechanics leads to the Navier-Stokes equations 
\eqref{ns}. It follows that this scaling operation is a symmetry of 
the same equations. This is easily directly verified. In particular, 
if the fields $v^i(x, t)$ and $p(x, t)$ obey the unforced Navier-Stokes
equations, then the rescaled fields $\epsilon v^i(\epsilon x, \epsilon^2 t)$ and 
$\epsilon^2 p(\epsilon x, \epsilon^2 t)$ also obey the same equations. 
Consequently, the scaling operation described above is a symmetry of the 
unforced Navier-Stokes equations.

As we have described above, the Navier-Stokes equations may be obtained 
as the scaling limit of any reasonable relativistic equations of fluid 
dynamics. In describing the connection with gravity, in most of the rest 
of this note, we will take the parent fluid dynamical theory to be 
conformal. It is natural to wonder how much of the full relativistic
conformal group descends to a symmetry of the Navier-Stokes equations,
and in what form it does so. It is obvious that the relativistic Poincare
group descends to the Galilean symmetry group of the Navier-Stokes equations.
In the next section we explain that the scaling symmetry operation 
described in the previous paragraph is loosely related to the dilatation
operator of the parent relativistic conformal theory. \footnote{This loose `descent' is 
analogous to the relation of the dilatation operator of the Schroedinger group 
to the generator of scale transformations of the massless Klein Gordon 
equation.} Further we demonstrate that all spatial special conformal 
transformations also descend to exact symmetries of the Navier-Stokes equations.
After our scaling these transformations effectively turn out to be the `boost'
to a uniformly accelerated frame; the inertial forces one has to deal with
when working in a non-inertial frame are compensated for by a shift of the
pressure. These spatial special conformal transformations, together with the
Galilean group and the scaling symmetry described above, form the $(d+2)(d+1)/2-1$ dimensional symmetry algebra of the Navier-Stokes equations. We list the
commutation relations of this algebra\footnote{The nonrelativistic conformal
symmetry group of the Navier-Stokes equations includes the contraction 
of  spatial conformal generators $K_i$ but does not include a generator 
that descends from the temporal conformal generator $K_0$. Our algebra is 
distinct from the Schroedinger group studied for example in \cite{Son:2008ye}} 
, and the action of its generators on the velocity fields, in detail in the 
next section. 

The conformal symmetry algebra described above is just subset of the full
infinte dimensonal symmetry algebra of the Navier Stokes equations 
\cite{Gusyatnikova} (see also \cite{O'Raifeartaigh:2000mp,Hassaine:2000ti
,Duval:2008jg} for other related work) . The 
additional generators of the full symmetry algebra are very easy to describe; 
they consist of boosts to a reference frame whose velocity is homogeneous 
in space but an arbitrary function of time. Just as in our discussion in 
the paragraph above, the pseudo force from such a frame change may be 
cancelled by an appropriate shift in pressure, and so is a symmetry of 
the Navier Stokes equations. \footnote{We thank J. Maldacena and R. Gopakumar 
for suggesting a symmetry enhancement along these lines, 
and thank J. Maldacena for drwaing our attention to \cite{Gusyatnikova}, see 
especially the top of pg 68. R. Gopakumar and collaborators are currently 
further studying this algebra and its extensions.}

The scaling limit described above admits an interesting generalization. Consider
the equations of fluid dynamics on a base manifold  $G_{\mu\nu}=g_{\mu\nu} +
H_{\mu\nu}$ where $H_{\mu\nu}$ is small. By taking all terms that depend on
$H_{\mu\nu}$ to the RHS, $\nabla_\mu T^{\mu \nu}=0$ reduces effectively to the equations fluid dynamics on the base space $g_{\mu \nu}$ forced by an $H_{\mu\nu}$ 
dependent forcing function. If we combine the scaling described in the paragraph above 
with the $H_{00}, ~ H_{ij}= {\cal O}(\epsilon^2)$ and $H_{i0}=\epsilon A_i(x^i, t)$,
the effective resultant forcing function survives and is finite in  the
$\epsilon \to 0$ limit. It turns out that the effective scaled forcing function
depends only on  $A_i(x^i, t)$ and has a very simple form. It is precisely the
force applied on a charged fluid by effectively a background electromagnetic 
potential $A_0=0,~A_i=A_i(x^i, t)$. Consequently the `magnetohydrodynamical'
Navier-Stokes equations (i.e. the Navier-Stokes equations with a forcing 
function from an arbitrary background electromagnetic field) follows as a
universal result of a scaling of the equations of relativistic hydrodynamics
with small metric fluctuations. 

We now return to the duality between gravity and fluid dynamics. 
We apply the scaling limit described in the previous paragraphs to the 
equations of fluid dynamics that are holographically dual to gravity. 
This procedure gives us an asymptotically locally AdS$_{d+1}$ 
gravity dual to any solutions of the magnetohydrodynamical 
Navier-Stokes equations. The resultant metric describes small - but non linearly
propagating fluctuations about a uniform black brane. We present the explicit 
form of the resultant bulk metric in section \ref{sec:gravitydual} below.

It follows from our discussion that every solution to the Navier-Stokes 
equations \eqref{ns} is also a scaling limit of a solution to Einstein's 
equations with a negative cosmological constant with one important caveat. 
The proviso 
is that many actual solutions of the Navier-Stokes equations describe fluids
subject to hard wall type boundary conditions, and we do not (yet?) 
understand how to generate gravitational duals of these boundary conditions. 
However the qualitative effects of boundary conditions may easily be 
 mimicked by appropriate forcing functions which are completely in our hands. 
In particular, while several experiments that study fully developed steady
state turbulence do so in fluids with hard wall boundary conditions,
there should be no barrier to setting up the same phenomenon for a fluid 
with no boundaries (e.g. on $R^{d}$ or on a compact manifold) with the 
appropriate forcing function. In order to make this 
expectation concrete we have identified one forcing function 
(of a likely infinite plethora of possibilities) applied to a fluid on 
$R^d$, whose steady state end flow we expect to be turbulent at asymptotically 
high Reynolds numbers. In the rest of this introduction we describe 
this forcing function, and the reason we expect the flow it generates to 
be turbulent.  

Consider the forcing function $A_x= \alpha e^{i \omega_0 t + i k_o y}
+ $ cc, together with $A_i=0$ for $i\neq x$ acting on (for concreteness) 
a fluid, on $R^{2,1}$ whose spatial sections are parameterized by the 
Cartesian coordinates $x$ and $y$. This gauge fields sets up a time and 
$y$ dependent electric field in the $x$ direction, together with a magnetic 
field in the plane. It is very easy to find one exact solution to the 
equations of fluid dynamics subject to this forcing function. On this 
solution $v_x$ is proportional to $A_x$ and $v_y=0$. This solutions 
describes a fluid driven into motion in the direction of an applied electric 
field (the $x$ direction), while Lorentz forces from the magnetic field,
in the $y$ direction,  are balanced by pressure gradients. All nonlinear
terms in the Navier-Stokes equations vanish when evaluated on our solution.
However nonlinear terms have an important effect on the dynamics of small
fluctuations about our solution at high Reynolds number. In particular, 
in  Appendix \ref{app:stab} below we demonstrate that nonlinear terms 
drive some fluctuation modes (with momentum in the $x$ direction) 
unstable at high enough Reynolds numbers. In order to explain the possible
significance of this instability, it is useful to recall the usual 
situation with fluids at high Reynolds numbers. 

As we have described earlier in our introduction, fluid flows at high 
Reynolds numbers are very rich, and of potential theoretical interest.
Nonetheless, in every highly symmetrical situation, there exists a simple 
`laminar' solution, that preserves all the symmetries of the problem, 
at every value of the Reynolds number. This solution is often simple
to determine analytically, and certainly shows none of the fascinating
phenomenology of turbulence. However in interesting situations this
solution becomes `tachyonic' i.e. goes unstable to linear fluctuations
that break some of the symmetries of the problem, above a critical
Reynolds numbers. The `end point' of this `tachyon condensation' 
typically has richer dynamical behavior than the original
solution itself. As the Reynolds number is further increased, further 
instabilities are usually triggered, and at arbitrarily high
Reynolds numbers flow is turbulent.

We suspect that the dynamical pattern described in the previous paragraph
applies to the exact solution described in this paper. As we currently 
have no theoretical tools to predict the onset of turbulence in any fluid flow 
this is necessarily a guess, but one that we believe is natural, given the 
results of our stability analysis. This guess suggests that the stable steady 
state solution to AdS$_4$ gravity, with the effective gauge field $A_x$ 
described above,  is dual to a turbulent fluid flow at high Reynolds numbers. 
Of course the particular situation described above is only one of a plethora 
of possibilities. We describe this solution in detail in Section 
\ref{sec:simplesteady} and Appendix \ref{app:stab} below only in order 
to have one concrete example of a gravitational set up that is likely to
be dual to a turbulent fluid flow; not because we think that our particular
is distinguished in any way.

We believe it is likely that the fluid - gravity map will lead to interesting
new insights on the nature of solutions of Einstein gravity in the presence 
of a horizon\footnote{For instance, consider a black 
brane in asymptotically flat space. At sufficiently low energies, 
a beam of gravitons shot at this brane perturbs the non normalizable 
boundary conditions of the effectively AdS near horizon region of the 
brane. It may be possible to choose this perturbation to drive the 
gravity in the near horizon AdS region turbulent.}. It also does not 
seem impossible that gravitational techniques and methods will prove useful in 
bringing new insights into the investigation of fascinating fluid phenomena 
like turbulence. In particular, it would be fruitful to understand the 
Kolmogorov laws on well-developed turbulence and their modification 
within the gravity framework. The symmetry algebra \eqref{comrel} may also
throw new light on these issues. We leave further investigation of these
issues to future work.

\textbf{Note Added} : While we were completing this paper we received the 
preprint \cite{Fouxon:2008tb} which has substantial overlap with 
Section 2 of this paper.

\section{Scaling the Navier-Stokes Equation}

\subsection{The Navier-Stoke Equation as a universal limit of fluid dynamics}

In this section we will display a scaling limit \footnote{This section was 
worked out in collaboration with J. Maldacena.} that reduces 
fluid dynamical equations of the form $\nabla_\mu T^{\mu\nu}=0$, to the 
incompressible non-relativistic Navier-Stokes equations usually studied in 
fluid dynamics text books (see e.g. \cite{1959flme.book.....L}).

The equations of relativistic fluid dynamics are 
\begin{equation}\label{rns}
\begin{split}
\nabla_\mu T^{\mu \nu}& =0 \\
T^{\mu \nu}&= \rho u^\mu u^\nu + P {\cal P}^{\mu\nu}
-2 \eta \sigma^{\mu\nu} -\zeta \theta {\cal P}^{\mu\nu} + \ldots\\ 
{\cal P}^{\mu\nu}&=g^{\mu \nu}+ u^\mu u^\nu, ~~~\\
\sigma^{\mu\nu}&={\cal P}^{\mu \alpha} {\cal P}^{\nu \beta} 
\left( \nabla_ \alpha u_\beta + \nabla_\beta u_\alpha
- \frac{g_{\alpha \beta}}{d-1} \partial. u \right), 
~~~\theta= \nabla^\beta  u_\beta 
\end{split}
\end{equation}
Here $P$ is the pressure, $\rho$ the energy density, $\eta$ the shear 
viscosity, $\zeta$ the bulk viscosity of the fluid, and $g_{\alpha \beta}$ 
the metric of the space on which the fluid propagates. Each of the fluid 
quantities listed above may be regarded as a function of the fluid 
temperature - and also of chemical potentials if the fluid is charged. 
The $\ldots$ in the equation above refer to terms of second or higher
order in spacetime derivatives. 

If the fluid is charged, then we must supplement the equation \eqref{rns}
with an equation of charge conservation for every conserved charge. 
Below, we comment briefly on the scaling limit of this equation. 

\subsection{A scaling limit}

Now let us study the motion of a fluid on a metric of the form  
$G_{\mu\nu}=g_{\mu\nu}+H_{\mu\nu}$. We assume that the background 
metric $g_{\mu\nu}$ has the form  
\begin{equation} \label{metform}
g_{\mu\nu}dx^\mu dx^\nu= -dt^2 + g_{ij}dx^i dx^j.
\end{equation}
(this is simply a choice of coordinate system, for a large class of metrics)
while the small fluctuation $H_{\mu\nu}$ is completely arbitrary. 
As we have explained in the introduction, we intend to view the fluid 
flow on the space with metric $G_{\mu\nu}$ as an effectively forced flow 
on the space with metric $g_{\mu \nu}$. In order to do this we need 
to map the velocity field ${\tilde u}^\mu$ on the space $G_{\mu\nu}$ to 
a velocity field on $g_{\mu\nu}$. This map, which must respect the 
requirement that $u^2={\tilde u}^2 =-1$, may be chosen in a natural fashion 
if one thinks of the velocity field as generated by the path of `particles' 
through spacetime. We now describe this in more detail. 

Let the fluid velocity on the space $G_{\mu\nu}$ be given by 
${\tilde u}^\mu$ where 
\begin{equation} \label{fluvel1}
{\tilde u}^\mu  = \frac{1}{\sqrt{V^2}} \left( 1, {\vec V} \right)
\end{equation}
where ${\vec V}$ is a $d-1$ spatial vector with components $V^i$, 
$V^\alpha$ is a $d$ component object with components $(1, {\vec V} )$ and 
$V^2$ is $G_{\alpha \beta} V^\alpha V^\beta$ (the indices $\alpha, \beta$ run 
over $d$ spacetime indices while the indices $i, j$ run over the $d-1$ spatial
indices). Expanding ${\tilde u}^\mu$ 
to first order in $H_{\alpha \beta}$ we have 
\begin{equation} \label{fluvel}
\begin{split} 
{\tilde u}^\mu& = u^\mu + \delta u^\mu + \ldots \\ 
u^\mu & = \frac{1}{\sqrt{1 - g_{ij} V^i V^j}} \left( 1, {\vec V} \right)\\
\delta u^\mu & = -u^\mu \frac{u^\alpha u^\beta H_{\alpha \beta}}{2}\\ 
\end{split}
\end{equation}
where $u^\mu$ is the $d$ velocity of the fluid referred to the metric 
$g_{\mu\nu}$. All terms in $\delta u^\mu$ above depend on the fluctuation
metric $H_{\alpha \beta}$; below we will take all these terms to the RHS of 
\eqref{ns} and view them as contributions to the effective forcing function.

One obvious solution to the equations of fluid dynamics on the space 
with metric $g_{\mu\nu}$ is simply the fluid at rest with constant 
pressure $P_0$ and density $\rho_0$. We now turn to a particular kind of 
small amplitude and long 
distance fluctuation about this uniform fluid at rest and with pressure $P_0$
and energy density $\rho_0$  on a manifold that is `close' to $g_{\mu\nu}$. 
More specifically we set 
\begin{equation} \label{scalingn}
\begin{split}
H_{00}& = \epsilon^2 h_{00}(\epsilon x^i, \epsilon^2 t)\\
H_{0i}& = \epsilon A_i (\epsilon x^i, \epsilon^2 t)\\
H_{ij}&= \epsilon^2 h_{ij} (\epsilon x^i, \epsilon^2 t)\\
V^i&=\epsilon v^i(\epsilon x^i, \epsilon^2 t)\\
\frac{P-P_0}{\rho_0 + P_0}&= \epsilon^2 p(\epsilon x^i, \epsilon^2 t) 
\end{split}
\end{equation}
and take $\epsilon$ to be arbitrarily small. Although we have explicitly 
listed only the scaling of the pressure $P$ above, 
the energy density $\rho$ and viscosity $\nu$ also scale in a similar fashion 
(this is consistent with the fact that they are all functions of the 
same underlying variables). We have normalized pressure fluctuations by 
$\rho_0 + P_0$ rather than $P_0$ for future convenience.

We will now examine what happens to the Navier-Stokes equations under 
this scaling. Let us first start with the $0$ or temporal 
component of these equations. 
It is easy to check that 
\begin{equation}\label{zeroc} 
\begin{split}
\nabla_\mu T^{\mu 0}& = \epsilon^2 \left[\rho_e \left( \nabla_i v^i\right)
\right] + 
{\cal O}(\epsilon^4)\\
\rho_e&= \rho_0+P_0\\
\end{split}
\end{equation}
Consequently, in the limit of small $\epsilon$ this equation reduces simply 
to $\nabla_i v^i=0$, where $\nabla_i$ is the covariant derivative with 
respect to the purely spatial metric $g_{ij}$. 

Let us now turn to the spatial Navier-Stokes equations. After some calculation
we find 
\begin{equation}\label{ic1}
\begin{split} 
\nabla_\mu T^{\mu i}& = \epsilon^3 
\left[ \rho_e \nabla^i p + \rho_e \nabla_\mu \left( 
v^i v^\mu \right) -2 \eta \nabla_j 
\left(\frac{ \nabla^j v^i+ \nabla^i v^j}{2}-g^{ij}\frac{\vec{\nabla}. 
{\vec v}} {d-1} \right) -\zeta \nabla_i {\vec \nabla}. {\vec v}  -f^i \right]
+ {\cal O}(\epsilon^5)  \\
f^i&=\rho_e\left( \frac{\partial_i h_{00}}{2} - \partial_0 A_i 
- \frac{\partial_j (\sqrt{g} A_i v^j)}{\sqrt{g}} + v^j\partial^i A_j 
\right)
\end{split}
\end{equation}
where $v^\mu=(1, v^i)$. Using the equation $\nabla_i v^i=
\frac{\partial_i(\sqrt{g} v^i)}{\sqrt{g}}=0$, the coefficient of 
$\epsilon^3$ in the equation above may be simplified to 
\begin{equation}\label{ic}
 \nabla^i p + \partial_0 v^i 
+ {\vec v}. \nabla v^i - \nu \left(  \nabla^2 v^i  +   R^i_j v^j 
\right) 
= \frac{\partial^i h_{00}}{2} - \partial_0 A^i
+ F^i_j v^j
\end{equation}
where  $F_{ij}=\partial_i A_j-\partial_j A_i$ is the field strength 
for the vector field $A_i$ and $\nu=\eta/\rho_e$ is the 
`kinematical viscosity' of the fluid. 
\footnote{ Note also that we have used conventions in which
$$[\nabla_\rho, \nabla_\sigma]A_\nu=A_\beta R^\beta_{\nu\rho \sigma}, ~~~
R^\beta_{\alpha \theta \beta}= R_{\alpha \theta}$$ 
With these conventions the scalar curvature of a unit $d$ sphere is 
$d(d-1)$ and the Ricci tensor of a unit sphere is given by 
$R_{ij}=(d-1)g_{ij}$. }
 \eqref{ic} takes a somewhat simpler form in terms of slightly redefined 
variables. Let $$A_i= a_i +\nabla_i \chi$$
where $\chi$ is chosen to ensure that $\nabla_i a^i=0$. This is the usual 
split of a gauge field into its pure curl and pure divergence parts. Note 
that 
$$f_{ij}\equiv \partial_i a_j-\partial_j a_i= F_{ij}.$$
We also define the effective pressure 
$$p_e=p-\frac{1}{2}  h_{00} + {\dot \chi} $$
in terms of which \eqref{ic} reduces to  
\begin{equation}\label{ics}
 \nabla_i p_e + \partial_0 v_i 
+ {\vec v}. \nabla v_i - \nu \left( \nabla^2 v_i +  R_{ij} v^j \right)  
= -\partial_0 a^i
- v^j f_{ji}
\end{equation}
\eqref{ics} is precisely the Navier-Stokes equation with forcing function 
generated by an effective background electromagnetic field (with $a^0=0$ and 
spatial vector $a^i$) on the effectively charged fluid.  

\subsection{Charged Fluids}

If the fluid under study carries an extra set of conserved charges, it 
obeys an extra set of conservation equations of the form 
$\nabla_\mu J^\mu=0$. 
It is easy to verify that these equations all reduce to the condition 
that the fluid velocity is divergence free in the scaling limit under 
study in this section. Consequently charged fluids obey 
the same equations as uncharged fluids in the scaling limit under study 
in this subsection.

\subsection{Reduction of Cauchy Data}

Returning to the study of uncharged fluids, the Cauchy data\footnote{We 
thank S. Trivedi for very useful discussions on the topic of this subsection.}
of the 
parent relativistic fluid dynamical equations consists 
of $d$ real functions of space; the value of the pressure field and the 
value of the $d-1$ independent velocity fields on an initial time slice. 
As the fluid dynamical equations 
are of first order in time, the Cauchy data of the problem does not 
include the time derivatives of all these fields. 

Now let us examine the Cauchy data of the incompressible Navier-Stokes 
equations. Note that the spatial divergence of \eqref{ic} 
\begin{equation}\label{press}
\nabla^2 p_e = -\nabla_i v^j \nabla_j v^i - v^i v^j R_{ij}  \
+ \nabla_i \left[ \left( -\nu R^i_j + f^i_j \right) v^j \right] 
\end{equation}
determines the pressure of the fluid as a function of the fluid velocity 
(only the velocity - not its time derivatives ).  In other words the independent data 
in the fluid is simply given by $d-2$ real functions that parameterize 
an arbitrary divergence free velocity field. 

It follows that two of the degrees of freedom of the equations of relativistic
fluid dynamics are lost on taking the scaling limit of the previous 
subsection. These two degrees of freedom  are simply the fluctuations 
of the pressure and the divergence of the velocity. At the linearized 
level these two degrees of freedom combine together in sound mode 
fluctuations. (Note that the relativistic dispersion of sound implies that 
a sound mode has twice as much data as a shear mode.). Consequently 
the reduction of Cauchy data in the scaling limit of this paper follows 
simply a nonlinear restatement of the observation that sound waves are 
projected out in our scaling limit.

\section{Symmetries}

As we have noted above, the Navier-Stokes equations may be obtained 
by applying the appropriate scalings to the equations of relativistic 
hydrodynamics with an arbitrary equation of state. The parent hydrodynamical 
system may in particular be chosen to be conformally invariant. 
It is natural to 
wonder whether either the dilatation or special conformal transformations of 
the parent theory descend to a symmetry of the the Navier-Stokes equations. 
In this section we answer this question in the affirmative. \footnote{This 
section was worked out in collaboration with R. Loganayagam.}

\subsection{Dilatations}

Let us first explain how this works for dilatations. Dilatations consist of 
a diffeomorphism $(x')^\mu=\frac{x^\mu}{\lambda}$, $T'=T$, $(u')^\mu=
\frac{u^\mu}{\lambda}$ $(g')_{\mu\nu}=  \lambda^2 g_{\mu\nu}$ 
compounded with the Weyl transformation 
${\tilde x}^\mu= (x')^\mu$, ${\tilde g}_{\mu\nu}= \frac{(g')_{\mu\nu}}
{\lambda^2}$, ${\tilde u}^{\mu}= \lambda (u')^{\mu}$, 
${\tilde T}= \lambda T $. In sum  
$${\tilde x}^\mu = \frac{x^\mu}{\lambda}, ~~~\tilde{g}_{\mu\nu}= g_{\mu\nu}, ~~~ \tilde{u}^\mu= u^\mu, ~~~\tilde{T}=\lambda T .$$ 
This action on coordinates, temperatures and fields, is a symmetry of the 
equations of conformal relativistic fluid dynamics. While this operation 
commutes with our scaling limit, it is not by itself a symmetry of the 
Navier-Stokes equations. This is because the kinematical viscosity 
$\nu$, regarded as a parameter in the Navier-Stokes equations, is proportional 
to $\frac{1}{T}$ and so changes under the variable transformation described 
above. The Navier-Stokes equations are however  
simple enough to allow dilatation transformations listed above to be 
modified into a true symmetry of the Navier-Stokes equations 
by `absorbing' the transformation of $\nu$ into `anomalous' transformations
of time and velocity. The result of this procedure is simply \eqref{scaling}
with which we started this note. 

\subsection{Special Conformal Transformations}

The situation is more straightforward for special conformal transformations. 
The scaling law for the velocity and temperature fields, under a special 
conformal transformation, may also be obtained by compounding  a 
diffeomorphism with the appropriate Weyl transformation. Restricting to 
infinitesimal conformal transformations, we find \footnote{In order to verify the covariance of local equations under these 
symmetry transformations, it is sufficient to omit the terms proportional 
to $\delta x^\mu \partial_\mu$ but instead to transform all derivatives 
according to the rule $$\delta \left( \partial_\beta \right) = 2 \left[ c_\beta  x.\partial -x_\beta c.\partial + x.c \partial_\beta \right]$$}
\footnote{Under this transformation, the shift in a conformal stress tensor is given by 
\begin{equation}\label{transstress} 
\delta T^{\mu\nu}= 2 d (c.x) T^{\mu\nu} + 2 (x^\lambda c^\mu -x^\mu c^\lambda) 
T^\nu_\lambda +  2 (x^\lambda c^\nu -x^\nu c^\lambda)T^\mu_\lambda
- \delta x^\lambda \partial_\lambda T^{\mu \nu} \end{equation}
Upon accounting also for the shift in the derivatives, 
it is easy to convince oneselves that the equation of energy momentum 
conservation, for any identically traceless stress tensor, 
is invariant under special conformal transformations. } 
\begin{equation}\label{conftransfrel} \begin{split}
\delta x^\mu& = -2 c.x x^\mu +x^2 c^\mu \\
\delta u^\mu&= -2\left[ x^\mu c^\nu-x^\nu c^\mu\right] u_\nu
- \delta x^\nu \partial_\nu u^\mu  \\
\delta T&= 2 c.x T - \delta x^\nu \partial_\nu T \\
\end{split}
\end{equation}

Now note that special conformal transformations induce an 
additive shift on the temperature fluctuation, $\delta T$, proportional 
to $x.c T_0$ where $T_0$ is the temperature of the background. In order 
that this shift respect the $\epsilon^2$ scaling of $\delta T$ we are 
required to scale $c_0 \propto \epsilon^4$ and $c_i \propto \epsilon^3$.
Imposing this scaling and retaining terms only to leading order in the 
velocity expansion \eqref{conftransf} reduces to\footnote{As above, in verifying the covariance of equations under 
these transformation, it is sufficient to omit the terms proportional
to $t^2 c_j \partial_j$ above, but instead to transform derivatives 
according to  $$\delta \left( \partial_t \right) = 2 t c^i \partial_i ~~~
\delta \left( \partial_i \right)  =0.$$} 
\begin{equation}\label{conftransf} \begin{split}
\delta t & = 0  \\
\delta x^i& = - t^2 c^i \\
\delta v^i &= -2 c^i t  + t^2  c_j \partial_j v^i  \\
\delta T&= 2 (-c^0 t + c^i x^i)  T  + t^2 c_j \partial_j T\\
\end{split}
\end{equation}

It is clear that the symmetry generated by $c^0$ acts trivially (it does 
not act on coordinates or velocities, but merely generates a shift, linear 
in time, of the pressure; more about this below). However the symmetries generated by 
$c^i$ act nontrivially, 
and are directly verified to be  symmetries of the Navier-Stokes 
equations. \footnote{Under the transformations listed in \eqref{conftransf} 
we have $\delta p_e= \frac{\delta p}{d P_0}= 2 c.x $ so that 
$\delta \partial_i p_e= 2 c^i$. Further $\delta( {\dot v^i} + v.\nabla v^i )
= -2 c^i$ and the viscous term is unchanged. Adding all terms together we 
have a symmetry of the equations.}

\subsection{The Full Symmetry Group}

We have thus discovered that the Navier-Stokes equations enjoy invariance
under a conformal symmetry group. The generators of this group are the dilatation $D$, 
special conformal symmetries $K_i$ Gallilian boosts $B_i$, the generator 
of time translations (energy) $H$, momenta $P_i$ and spatial rotations 
$M_{ij}$. The action of these generators on velocity fields is given by 
\begin{equation}\label{gentr} \begin{split}
D v^j  =& \left( -2 t \partial_t - x^m \partial_m - 1 \right) v^j  \\
K_i v^j=& -2 t \delta_{ij}  + t^2  \partial_i v^j \\
B_i v^j= & \delta_{ij} -t \partial_i v^j\\
H v^j = & - \partial_t v^j \\
P_i v^j=& - \partial_i v^j \\
M_{ik} v^j=& \delta_{ij} v^k - \delta_{kj} v^i - (x^k \partial_i - x^i 
\partial_k) v^j \\
\end{split}
\end{equation}

The commutation relations between these various generators is given by 
\begin{equation}\label{comrel} \begin{split}
[D, K_i]=& -3 K_i\\
[D, B_i]=& -B_i \\
[D, H]=& 2 H\\
[D, P_i]=& P_i  \\
[D, M_{ij}]=&0  \\
[M_{ij}, P_k]=& -\delta_{ik} P_j + \delta_{jk} P_i \\
[ M_{ij}, K_k]= &-\delta_{ik} K_j + \delta_{jk} K_i \\
[ M_{ij}, B_k]= &-\delta_{ik} B_j + \delta_{jk} B_i \\
[M_{ij}, H]=&0\\
[K_i, P_j]=& 0\\
[K_i, B_j]=& 0 \\
[K_i, H]=& -2 B_i\\
[H, B_j]=& -P_i\\
\end{split}
\end{equation}

The symmetry algebra listed in \eqref{comrel} may presumably be obtained from the appropriate
contraction of the parent symmetry algebra $SO(d,2)$. A related study is currently 
under progress in \cite{gopakumaretal}. 

In addition to the symmetries listed above, the Navier-Stokes equations 
have an infinite dimensional group of trivial symmetries, under which 
the pressure is simply shifted by an arbitrary function of time. These 
are symmetries of the equation because only gradients of the pressure 
enter the Navier-Stokes equations, and they are trivial because the 
pressure is not really an independent variable of the Navier-Stokes equations, 
which may in fact be eliminated by taking the curl of those equations. 
For this reason one need not keep track of the action of symmetry generators
on the pressure. However it is not difficult to do so and we find 

\begin{equation}\label{gentrp} \begin{split}
D p_e  =& \left( -2 t \partial_t - x^m \partial_m - 2 \right) p_e  \\
K_i p_e =&  2 x^i  + t^2  \partial_i p_e \\
B_i p_e = & -t \partial_i p_e\\
H p_e = & - \partial_t p_e \\
P_i p_e =& - \partial_i p_e \\
M_{ik} p_e =& - (x^k \partial_i - x^i 
\partial_k) p_e \\
\end{split}
\end{equation}
This action of the symmetry generators on the pressure field 
do not quite yield the commutation relations \eqref{comrel}, but instead have 
additional terms on the RHS corresponding to generators of the trivial 
symmetries referred to above (i.e. generators that shift the 
pressure field by a function of time). Spatial derivatives of the pressure field, 
however, honestly transform according to the algebra \eqref{comrel}. Consequently,
the symmetry algebra \eqref{comrel} is not represented on the pressure field 
itself, but on its spatial derivatives.  This is vaguely 
reminiscent of the fact that the two dimensional conformal group 
has well defined action on all derivatives of massless scalar fields, but not 
on the field itself.

\section{The gravity solution dual to fluid dynamics in the scaling regime}\label{sec:gravitydual}

\subsection{The dual bulk metric}

As we have described above, any solution of the incompressible non-relativistic
Navier-Stokes equations solves the equations of fluid dynamics 
dual to gravity up to ${\cal O}(\epsilon^3)$, under the scaling listed 
in the previous section.  Now in \cite{ Bhattacharyya:2008ji, 
Haack:2008cp, Bhattacharyya:2008mz} the equations 
of fluid dynamics are obtained as the  Einstein constraint 
equation of a bulk asymptotically locally AdS$_{d+1}$ space. It is thus 
natural to define the gravitational dual to a 
solution of the Navier-Stokes equations as any small fluctuation about 
a black brane background that solves all of Einstein's equations (constraint
as well as dynamical) to cubic order in $\epsilon$. Adopting this definition, 
it is easy to read off the bulk metric dual 
to the Navier-Stokes equations from an appropriate 
scaling of the bulk metric of \cite{ Bhattacharyya:2008ji, 
Haack:2008cp, Bhattacharyya:2008mz}. 

In computing the bulk metric upto ${\cal O}(\epsilon^3)$ in the sense described above, it turns out that terms from only the zeroth order and the first 
order in the derivative expansion of the gravitational solutions of 
\cite{ Bhattacharyya:2008ji, Haack:2008cp, Bhattacharyya:2008mz} are relevant.
The metric up to first order in derivative has the following form
\begin{equation}
\begin{split}
ds^2  &= ds_0^2 + ds_1^2\\
\text{where}&\\
ds_0^2 &= -2 u_\mu dx^\mu dr + \frac{1}{b^d r^{d-2}}u_\mu u_\nu dx^\mu dx^\nu + r^2 g_{\mu\nu}dx^\mu dx^\nu\\
ds_1^2 &= -2 r u_\nu \left(u^\alpha\bar\nabla_\alpha\right) u_\mu dx^\mu dx^\nu + \frac{2}{d-1}r \left(\bar\nabla_\alpha u^\alpha\right) u_\mu u_\nu dx^\mu dx^\nu + 2 b r^2 F\left(b r\right) \sigma_{\mu\nu}dx^\mu dx^\nu\\
\sigma_{\mu\nu} &= \frac{1}{2}\left(\bar\nabla_\mu u_ \nu + \bar\nabla_\nu u_ \mu\right) + \frac{1}{2}\left(u_\nu \left(u^\alpha\bar\nabla_\alpha\right) u_\mu + u_\mu \left(u^\alpha\bar\nabla_\alpha\right) u_\nu\right) - \frac{1}{d-1}\left(\bar\nabla_\alpha u^\alpha\right)\left(u_\mu u_\nu + g_{\mu\nu}\right)\\
F(x) &= \int_{x}^\infty dy \left(\frac{y^{d-1} - 1}{y(y^d -1)}\right)\\
b &= \frac{d}{4 \pi T} = b_0 + \delta b\\
T &= T_0 + \delta T
\end{split}
\end{equation}
Here $\bar\nabla$ denotes the covariant derivative with respect to the full boundary metric which is equal to a background $g_{\mu\nu}$ plus perturbation $H_{\mu\nu}$. $T_0$ is the temperature of the background blackbrane.
The terms that appear in the metric involve covariant derivatives of the $d$ velocity $u_\mu$ with respect to the full boundary metric.  These terms can be expressed as covariant derivatives of the $d-1$ velocity $v_i$ and the metric perturbation $A_i = H_{0i}$ with respect to spatial part of the background metric $g_{ij}$.  

\begin{equation}\label{deri}
\begin{split}
\bar\nabla_iu_j &=\nabla_iv_j + {\cal O}(\epsilon^4)\\
\bar\nabla_iu_0 + \bar\nabla_0u_i &= \partial_0(v_i + A_i) - \frac{1}{2}\partial_i h_{00} - \frac{1}{2}\partial_i(v_jv^j) - v^jF_{ij} + {\cal O}(\epsilon^4)\\
\bar\nabla_\mu u^\mu &= \nabla_jv^j + {\cal O}(\epsilon^4)\\
u^\mu\bar\nabla_\mu u_0 &= {\cal O}(\epsilon^4)\\
u^\mu\bar\nabla_\mu u_i &= \partial_0(v_i + A_i) - \frac{1}{2}\partial_i h_{00} + (v^j\nabla_j)v_i - v^jF_{ij} + {\cal O}(\epsilon^4)\\
F_{ij} &= \partial_i A_j - \partial_j A_i
\end{split}
\end{equation}
 Here $\nabla$ denotes the covariant derivative with respect to $g_{ij}$. The raising and lowering of the ${i,j}$ indices are also with respect to the metric $g_{ij}$. To simplify the expression of $\sigma_{\mu\nu}$ in \eqref{deri} the constraint $\nabla_i v^i = 0$ has been used.
Using these expressions the derivative part of the metric can be written as
\begin{equation} \label{fres}
\begin{split}
ds_1^2 &= b_0 r^2 F(b_0 r)\left(\nabla_iv_j + \nabla_j v_i\right)dx^i dx^j - 2 b_0 r^2 F(b_0 r)v^j\left(\nabla_iv_j + \nabla_j v_i\right)dt~dx^i\\
& +2r \left(\partial_0(v_i + A_i) - \frac{1}{2}\partial_i h_{00} - v^j F_{ij} + (v^j\nabla_j)v_i\right)dt~dx^i + {\cal O}(\epsilon^4)\\
\end{split}
\end{equation}
Here the first term is of order $\epsilon^2$ and the last two terms are of order $\epsilon^3$. Since from the constraint equations $\nabla_i v^i = 0$, there is no contribution from the scalar sector. 
The zeroth order metric can also be expanded in powers of $\epsilon$. It turns out that to solve Einstein equation up to order $\epsilon^3$ ,it is sufficient to expand the zeroth order metric up to order $\epsilon^2$ in the fluctuations.
\begin{equation} \label{zres}
\begin{split}
ds_0^2 &= \frac{1}{b_0^d r^{d-2}} dt^2 + r^2\left(-dt^2 + g_{ij} dx^i dx^j\right)  + 2 dt~ dr\\
&- \frac{2}{b_0^dr^{d-2}}\left(A_i + v_i\right)dt~ dx^i - 2\left(A_i + v_i\right)dx^i dr\\
&+\frac{1}{b_0^{d+1}r^{d-2}}\left(-d~\delta b + v_jv^j - h_{00}\right)dt^2 + \frac{1}{r^{d-2}}\left(A_i + v_i\right)\left(A_j + v_j\right)dx^idx^j  - \left(-v_jv^j + h_{00}\right)dt~dr\\
\end{split}
\end{equation}
Here the first line is of order $\epsilon^0$, the second line is of order $\epsilon^1$ and the third is of order $\epsilon^2$. The full metric 
is given by the sum of \eqref{zres} and \eqref{fres}; and solves 
Einsteins equations to ${\cal O} (\epsilon^3)$ 
provided the velocity and temperature fields above obey the incompressible 
Navier-Stokes equations \eqref{ns}. These equations imply in particular that 
\begin{equation}\label{temp}
\frac{\nabla^2 T}{T_0} = -\nabla_i v^j \nabla_j v^i - v^i v^j R_{ij}  \
+ \nabla_i \left[ \left( -\nu R^i_j + F^i_j \right) v^j \right] +
\frac{1}{2} \nabla^2 h_{00} - \partial_0 (\nabla. A) 
\end{equation}
an equation that determines $\delta T$ (and hence $\delta b$ in \eqref{zres})
as a spatially nonlocal but temporally ultralocal functional of the velocity
fields $v^i$. It follows that though the  bulk metric at $x^\mu$ is 
determined locally as a function of temperatures and velocities at $x^\mu$, 
it is not determined
locally as a function of velocities at $x^\mu$. This non locality 
is a consequence of the infinite speed of sound (and consequent action at 
a distance) in our scaling limit.

\subsection{Comments on the bulk metric}

Several comments on the bulk metric \eqref{fres} and \eqref{zres} are in 
order. Let us first spell out some terminology. We refer to the parts of 
the bulk metric that are proportional to a linear combination of 
$dt^2$ or $\sum_i (dx^i)^2$ as its scalar components. Terms in the metric 
proportional to $dt dx^i$ are called its vector components, while terms 
proportional to $dx^i dx^j$ are referred to as its tensor components.

Now the derivative expansion of \cite{Bhattacharyya:2007jc,
Loganayagam:2008is,VanRaamsdonk:2008fp, Dutta:2008gf, Bhattacharyya:2008xc, 
Bhattacharyya:2008ji, Haack:2008cp,Erdmenger:2008rm,
Banerjee:2008th, Bhattacharyya:2008mz} solves the Einstein equations in 
a multi step process. In the first step the dynamical Einstein 
equations are used to determine the $r$ dependence of the bulk metric as 
a function of boundary data, in each of the sectors described above. 
The requirement of regularity of the future horizon is then used to 
 determine the boundary data of the tensor sector in terms 
of the boundary data in the vector and scalar sectors (roughly the fluid 
temperature and velocity). 

In the scaling limit described in the previous subsection, the tensor sector 
of the bulk metric is particularly simple. It is given by 
\begin{equation}\label{tensorsecmet}
 b_0 r^2 F(b_0 r)\left(\nabla_i v_j+\nabla_j v_i \right) dx^i dx^j
+ \frac{1}{r^{d-2}}\left(A_i + v_i\right)\left(A_j + v_j\right)dx^idx^j. 
\end{equation}
Note that even though the velocity $v_i$ is scaled to zero in the scaling 
limit under study in this paper, the leading order in $\epsilon$ 
metric \eqref{tensorsecmet} includes a quadratically nonlinear in the 
velocities. This is a consequence of the fact that we scale distances 
to infinity at the same rate as velocities to zero. Had we simply scaled 
amplitudes to be small, while keeping distance scales finite, we would 
have ended up with only the first of these two terms in \eqref{tensorsecmet}
and the resultant geometry would simply have been the dual of linearized 
fluid dynamics.  

The Navier-Stokes equations are obtained out of the Einstein constraint 
equations acting on the bulk metric described above. The first term in 
\eqref{tensorsecmet} gives rise to the viscous term in the Navier-Stokes 
equations, while the 
second term in \eqref{tensorsecmet} is the origin of the nonlinear 
convective term in those equations. It is consequently not surprising that 
ratio of the first to the second term in \eqref{tensorsecmet} is proportional 
to the Reynolds number $Re$ of the flow: in fact it is of order 
$ Re \times \frac{1}{r^d F(r)}$. Now $F(r) \sim \frac{1}{r}$.
at large $r$; so that the viscous term in \eqref{tensorsecmet}
appears to dominates the nonlinear term in that expression when 
$r > (Re)^{1/d-1}$. This appearance is atleast partly a 
coordinate artifact, as is evidenced by the fact that the two terms in 
\eqref{tensorsecmet} contribute equally to the 
Einstein constraint equations at every $r$ (this follows as as these 
equations are independent of $r$). \footnote{The ratio computed 
above naively suggests that nonlinearities 
at different effective Reynolds numbers are separated in the bulk radial 
coordinate; a result reminiscent of the approximate locality in scale 
of high Reynolds number fluid flows. While this interpretation is too quick for the purposes of 
evaluating the Einstein constraint equation,  perhaps the naive ratio 
of the two terms in \eqref{tensorsecmet} is 
physically relevant for some physically interesting 
questions. We feel this deserves further contemplation. }

\section{Simple Steady State Shear flows at arbitrary Reynolds Number}\label{sec:simplesteady}

Consider a fluid on $R^{d-1,1}$, subject to an effective forcing 
\begin{equation}\label{solga} \begin{split}
a_x& = \int d\omega d k  a(k, \omega) 
\exp \left[i k y +i \omega t \right] \\
a_i&=0, \  \  \  (i \neq x)
\end{split}
\end{equation}
Here $x$ and $y$ parameterize orthogonal displacements in the 
$d-1$ spatial directions, and $a(k, \omega)$ is any function. 

In the presence of this forcing function, it is easy to verify that the 
velocity field 
\begin{equation} \label{solvel} \begin{split}
v_x&=-\int dk d\omega \frac{a(k, \omega) e^{i k y +i \omega t}}{1- i 
\frac{\nu k^2}{\omega}}\\
v_i&=0 \   \   \  i \neq x 
\end{split}
\end{equation}
(together with the pressure field which may be obtained by integrating 
\eqref{press} ), is an exact solution of \eqref{ics}. 

In order to be concrete let us make the simple choice 
$a(k, \omega)= a(-k, -\omega)= \alpha \delta(k-k_0) \delta(\omega-\omega_0)$. 
In this special case the forcing function is given by 
\begin{equation}\label{solnt1} \begin{split}
a_x& =  \alpha \exp \left[i k_0 y +i \omega_0 t \right] +\  \  {\it cc} \\
a_i&=0, \  \  \  (i \neq x)
\end{split}
\end{equation}
and the velocity field is given by 
\begin{equation} \label{solspc} \begin{split}
v_x&=-\frac{\alpha e^{i k_0 y +i \omega_0 t}}{1-i 
\frac{\nu k_0^2}{\omega_0}} + \  \  {\it cc} \\
v_i&=0 \   \   \  i \neq x 
\end{split}
\end{equation}
\footnote{This special
solution may be thought of as living either on a spatial $R^{d-1}$ or 
on a spatial torus, whose $y$ edge is of size $\frac{2 \pi}{k_0}$. }
This solution is characterized by two dimensionless numbers,
$a=\frac{\alpha}{k_o \nu}$ and $b=\frac{\nu k^2}{\omega}$
The Reynolds number for this solution is given by 
\begin{equation}\label{re} R=\frac{vL}{\nu}= \frac{a}{\sqrt{1+b^2}}.
\end{equation}
and can be made large by taking $a$ to infinity at fixed $b$. 
This may be achieved, for instance, by increasing the strength 
of the forcing amplitude $\alpha$ keeping everything else fixed. 

In Appendix \ref{app:stab} we outline a linear stability analysis of the special 
flow described above. We find that in $d=2$ this flow is unstable at large 
enough Reynolds numbers for a range of the parameter $b$. As we have 
described in the introduction, this suggests that the flow is dynamically 
interesting, and likely turbulent, at asymptotically high Reynolds numbers. 
Assuming this is the case, we have identified atleast one gravitational 
system that is dual to steady state turbulence. In some detail, consider 
gravity with a negative cosmological constant, subject to the small $z$ 
boundary condition on the metric
$$ \lim_{z \to 0} ds^2=  
\frac{1}{z^2}\left( dz^2 -dt^2  + \epsilon dt dx^i a_i + dx^i dx_i 
\right)$$
(where $i= i \ldots d-1$ and $a_i$ is the gauge field \eqref{solga}). 
The steady state solution of this gravitational system is  
dual to a turbulent flow of $d$ dimensional fluid dynamics 
whenever the parameters in the function $a_i$ above are chosen to 
satisfy the inequality $a \gg {\sqrt{1+b^2}}$. 

In Appendix \ref{app:sphere} we have presented a generalization of the solution described 
in this section to the forced flow of a fluid on a sphere. In that situation
one should obtain the dual of a turbulent fluid in an spacetime that 
is asymptotically global AdS.

\subsection*{Acknowledgements}

We would like especially to thank J. Maldacena for discussions 
over which the scaling limit described in this paper were formulated,   
collaboration in the initial stages of this project, and useful discussions
and comments throughout its execution. We would also especially like to thank 
R. Loganayagam for collaboration on the calculations in Section 3, for 
useful discussions throughout the execution of this project and for 
several detailed comments on the manuscript. We have profited greatly from 
discussions with S. Trivedi and have also had fruitful discussions 
with J.Bhattacharya, R. Gopakumar, G. Mandal, 
H. Ooguri and  M. Van Raamsdonk. We would like to acknowledge 
useful discussion with the students in the TIFR theory room. 
The work of S.M. was supported in part by Swarnajayanti Fellowship. The
work of S.R.W was supported by J.C.Bose fellowship. We would 
all also like to acknowledge our debt to the people of India for their 
generous and steady support to research in the basic sciences.

%%%%%%%%%%%%%%%%%%%%%%%%%%%%%%%%%%%%%%%%%%%%%%%%%%%%%%%%%%%%%%%%%%%%%%%%%%
\section*{Appendices}
\appendix
%%%%%%%%%%%%%%%%%%%%%%%%%%%%%%%%%%%%%%%%%%%%%%%%%%%%%%%%%%%%%%%%%%%%%%%%%%

\section{Stability Analysis}\label{app:stab}

In this appendix we study linear fluctuations about the solution 
\eqref{solspc} for the special case $d=3$. In two spatial dimensions
the dual of a divergenceless velocity is the gradient of a scalar, and 
the field strength of a gauge field is itself dual to a scalar. In other 
words we can write 
\begin{equation}\label{redef} \begin{split}
v_i&= \epsilon_{ij} \partial_j \chi\\
f_{ij}&=\epsilon_{ij} f\\
\end{split}
\end{equation}

The Navier-Stokes equations \eqref{ics} may be rewritten (after eliminating 
the pressure by taking the curl) in terms of these scalars as
\begin{equation}\label{nsc}
-\nabla^2 {\dot \chi} + \nu \nabla^4 \chi - \epsilon^{jm} \nabla_m \chi 
\nabla_j \nabla^2 \chi + {\dot f} + \epsilon^{iq} \nabla_i \chi \nabla_q f
=0
\end{equation} 

In terms of these variables the solution \eqref{solspc} takes the form 
\begin{equation}\label{solnt} \begin{split}
f_0 &=  -i k_0 \alpha \exp \left[i k_0 y +i \omega_0 t \right] +\  \  {\rm cc} \\
\chi_0 & = - \frac{1}{i k_0} \times  \frac{\alpha e^{i k_0 y +i \omega_0 t}}{1-i 
\frac{\nu k_0^2}{\omega_0}} + \  \  {\rm cc} 
\end{split}
\end{equation}

\subsection{Setting up the Eigenvalue Equations}

Let us now set 
$$\chi= \chi_0 + \theta \left(   e^{i q_0 x} \int d\omega d k  
\chi_{q_0}(\omega, k) e^{i\omega y + i k x} 
+  e^{-i q_0 x} \int d\omega d k  
\chi^*_{q_0}(\omega, k) e^{-i\omega y - i k x} \right) $$
where $\theta$ is a small parameter, and solve \eqref{nsc} to linear order 
in $\theta$. Note that while our fluctuation has a specific frequency  
(namely $q_0$) in the $x$ direction, it is a sum over frequencies in the 
$y$ direction and in time. This is necessary as our background breaks 
translational invariance in the $y$ and $t$ directions, but preserves this 
invariance in the $x$ direction. It is easy to work out the linear equations
that our fluctuation coefficients $\chi_{q_0}(\omega, k)$ obey. We find 
\begin{equation}\label{recrel} \begin{split}
&\left[ i \omega(k^2+q_0^2) + \nu (k^2+q_0^2)^2 \right]
\chi_{q_0}(\omega, q) + 
\left[\beta (i q_0)\left( (k-k_0)^2+q_0^2-k_0^2 \right)
+ \alpha i q_0 k_0^2 \right]\chi_{q_0}(\omega-\omega_0, k-k_0) \\ &
+  \left[\beta^* (i q_0)\left( (k+k_0)^2+q_0^2-k_0^2 \right)
+ \alpha^* i q_0 k_0^2 \right]\chi_{q_0}(\omega+\omega_0, k+k_0)=0
\end{split}
\end{equation}
where 
$$\beta= - \frac{\alpha}{1+ \frac{\nu k_0^2}{i\omega_0}}.$$

Let us define 
$$f_k(\gamma, \kappa)= \chi_{q_0}\left[\omega_0(\gamma+k), k_0(\kappa+k) 
\right].$$
The fluctuation equation \eqref{recrel} may be recast in terms of $f_k$ 
as (we will usually omit to write the functional dependence of $f_k$ in the 
equations that follow)
\begin{equation}\label{recrelf}
\begin{split}
& \left[ \frac{i}{b}  (\gamma + k)((\kappa+k)^2 + c^2) + 
((\kappa+k)^2+c^2)^2 \right] f_k + i c a \left( -\frac{(\kappa+k -1)^2+ c^2-1}
{1-ib} +1 \right) f_{k-1} \\ & +  i c a^* \left( -\frac{(\kappa+k +1)^2+ c^2-1}
{1+ib} +1 \right) f_{k+1}
\end{split}
\end{equation}

The dimensionless quantities $a, b$ and $c$ are parameters in \eqref{recrelf} 
($a$ and $b$ were defined in section 4 while $c=\frac{q_0}{k_0}$).
$\gamma$ and $\kappa$ are variables in this equation;  
\eqref{recrelf} is the condition that a matrix acting on 
the columns $\{ f_k \}$ has a zero eigenvalue. Together with the 
physical requirement that $f_k$ decay at large $|k|$; this eigenvalue 
equation yields an expression for unknown temporal frequency $\gamma$ 
as a function of $\kappa$. \footnote{ Recall that $\kappa$ is a real number 
(in order that the mode in question is well defined at all $y$) of unit 
periodicity.}
That is, the equation 
\eqref{recrelf} should yield a dispersion relation of the form 
\begin{equation} \label{disprel}
\gamma= \gamma(\kappa, a, b, c, n)
\end{equation}
where the integer  $n$ labels which of the infinitely many solutions to the 
zero eigenvalue condition we have chosen to study. A result for $\gamma$ 
with a negative imaginary part represents an instability of the system. 

Let us study the large $|k|$ asymptotics of the variables $f_k$ in a little 
more detail. According to our boundary conditions, at large positive $k$
$f_{k+1}$ is much smaller than $f_{k-1}$. As a consequence \eqref{recrelf}
implies that 
\begin{equation}\label{ratpos}
\frac{f_k}{f_{k-1}}=\frac{i c a}{1-ib} \times 
\frac{k +2(\kappa -1)}{k^2(k+4 \kappa + \frac{i}{b})}  \times \left( 1 + 
{\cal O}(\frac{1}{k^2}) \right)
\end{equation}
It follows that, at large $k$, 
\begin{equation} \label{largeasymp}
f_k \approx D(a, b, c \kappa)  \left( \frac{i c a}{i-ib} \right)^k  
\frac{\Gamma(k+2\kappa -1)}{\Gamma(k+1)^2 \Gamma(k+ 4\kappa + \frac{i}{b} +1)}.
\end{equation}
where $D$ is a constant. This estimate is valid provided that $k \gg 1$ and that $k^2 \gg 
\frac{ca}{\sqrt{1+b^2}}$. Similarly, the behavior of $f_k$ at large negative values
of $k$ is given by 
\begin{equation}\label{smallasymp}
f_k \approx D'(a^*, b, c, \kappa) \left( \frac{i c a^*}{i+ib} \right)^{-k}  
\frac{\Gamma(-k-2\kappa -1)}{\Gamma(-k+1)^2 \Gamma(-k- 4\kappa + \frac{i}{b} +1)}
\end{equation}
where $D'$ is an independent constant. Note $\chi$ is real provided 
that $D(a, b, c, \kappa)=D'(a^*, b, -c, -\kappa)$; a condition that we 
consequently demand on physical grounds.  

We are interested in determining the dispersion relation \eqref{disprel} 
as the condition for the existence of a solution of \eqref{recrelf} that 
achieves both asymptotic conditions \eqref{largeasymp} and \eqref{smallasymp} 
(this leads to an equation as, generic solutions of \eqref{recrelf} that 
obey \eqref{largeasymp} blow up at large negative $k$).  

\subsection{Qualitative nature of the solution}

Let us first note that the off diagonal terms in \eqref{recrelf} are both 
proportional to $a \times c$. The proportionality to $a$ reflects the fact 
that the problem gets strongly coupled at high Reynolds numbers; we will 
have a lot more to say about this below. The fact that these terms are 
proportional to $c$, however, implies that fluctuations with no 
$x$ momentum are governed by the linear Navier-Stokes equations, and so 
are always stable. An instability, if it occurs, must necessarily 
break a symmetry (in this case of $x$ translations) that the background 
solution preserves.  

In much of the rest of this section we specialize our analysis to 
$\kappa=0$ and $c=1$ 
(i.e. $k_0=q_0$) but for a reasonably wide range of the parameters $a$ and 
$b$. We perform this specialization for convenience in the numerical analysis
that we will describe below; we expect qualitatively similar results at 
all values of $\kappa$ and reasonable nonzero values of $c$, though we have
not explicitly tested this expectation.
 
Let us start by painting a qualitative picture of physically relevant  
solutions of \eqref{recrelf}. We start with the simple limit 
$a \to 0$. Solutions are given exactly by 
$f_k= \delta_{k, K}$ with $\gamma=-K+ib (K^2+1)$ for $K$ ranging over 
integers. The time dependence of this solution is given by 
$f_k(t)=f_k(0)e^{-i K \omega_0 t - b \omega_o (K^2+1) t}$; it follows that 
all these modes decay in time, so that our solution is stable against small 
fluctuations.

At small but nonzero values of Reynolds number $Re=\frac{a}{\sqrt{1+b^2}}$
the solutions to \eqref{recrelf} are small perturbations of the solutions 
described in the paragraph above. On the $K^{th}$ such solution 
$f_k$ is nonzero at $k=K$, but decays rapidly as $k$ moves 
away from $K$. In particular $f_{k-K} \propto (Re)^{|k-K|}$. \footnote{This 
decay is visible, for instance, in the asymptotic forms \eqref{largeasymp} and 
\eqref{smallasymp}. }
The dispersion relation for $\gamma$, described in the previous paragraph, 
is corrected in a power series in $(Re)^2$. We will compute the first term 
in this series below. 

As the Reynolds number is increased, the width in the distribution of $f_k$
(as a function of $k$ centered around $k=K$) increases. For $Re \gg 1$ 
and $Re \gg \sqrt{n}$ this width is of order $k\sim (Re)^{\frac{1}{2}}$.\ 
(see \eqref{largeasymp} and \eqref{smallasymp}).

In summary, the fluctuation mode we study is  highly localized 
 about a particular $y$ momentum at small Reynolds number, but consists of 
a `cascade' or roughly equal superpositions of order $(Re)^{\frac{1}{2}}$
modes at large Reynolds numbers. The fluctuation modes oscillate rather 
than growing exponentially in time at small Reynolds numbers; however the 
stability properties of these modes must be elucidated by explicit calculation
at large Reynolds numbers. 

\subsection{Numerical Evaluation of the frequency}

In this subsection we present the results of our
rough numerical evaluation of the frequency $\gamma$ of the mode with $n=1$
as a function of the Reynolds number, at various values of $b$. The method 
we use is very simple. We start with a value of $k$ large enough for the 
asymptotic form \eqref{largeasymp} to be valid \footnote{We have performed 
computations with values of this starting $k$ ranging from 20 to 8. The plots 
presented below are generated using the starting value $10$. We have verified
that our starting value of $k$ is large enough, by checking that our results 
are not significantly affected by increasing $k$.} . 
We then use the recursion relations 
\eqref{recrelf} to evaluate $f_1/f_0$. We then independently evaluate the 
same ratio using \eqref{smallasymp} and recursion relations. Equating 
these expressions for the same ratio gives us an equation which we use to 
solve for $\gamma$ 
as a function of all other parameters. Of course this equation has several 
solutions. At small Reynolds number we choose the solution that is near 
to $\gamma=-1+ 2 ib$ (i.e. the mode with $K=1$ in the language of the previous 
subsection) , and then `follow' this root as we numerically increase 
the Reynolds number (in small steps) to large values. We have performed 
all our calculations using Mathematica. 

\begin{figure}[ht!]
 \begin{center}
\includegraphics[scale=0.9]{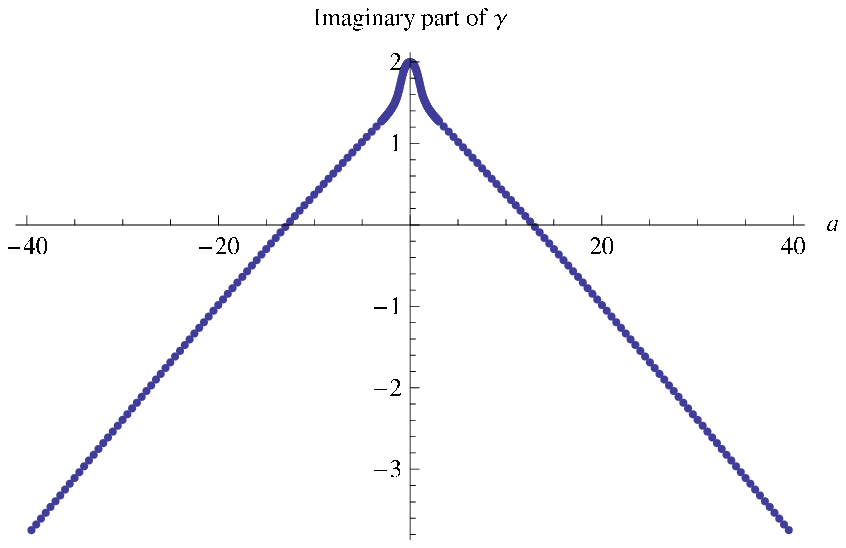}
\end{center}
\caption{The imaginary part of the frequency $\gamma$ plotted against $a$, 
for $\kappa=0$, $c=1$ and the root with $K=1$. Note that 
$Im (\gamma)=2$ at $b=0$ and decreases monotonically as $|b|$ is increased.}
\label{reg}
\end{figure}

\begin{figure}[ht!]
 \begin{center}
\includegraphics[scale=0.9]{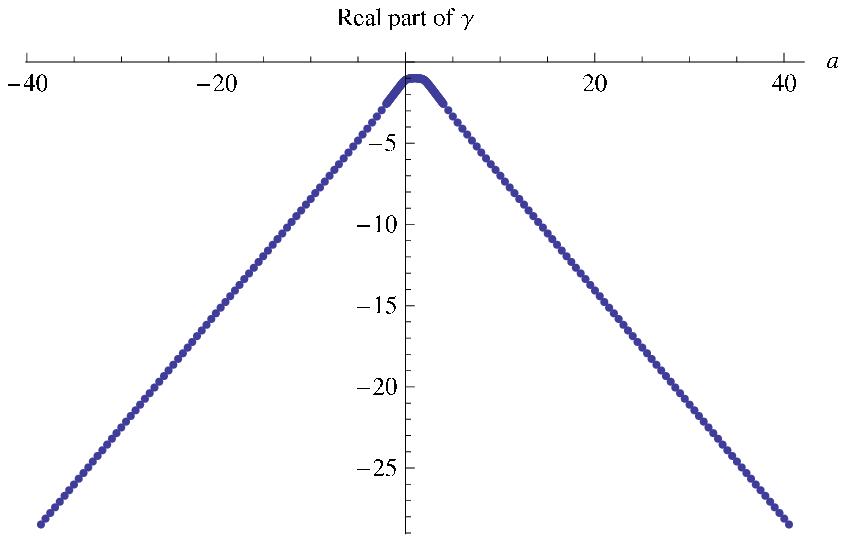}
\end{center}
\caption{The real part of the frequency $\gamma$ plotted against $a$, 
for $\kappa=0$, $c=1$ and the root with $K=1$. Note that 
$Re (\gamma)=-1$ at $b=0$ and decreases monotonically as $|b|$ is increased.}
\label{reis}
\end{figure}

Let us describe our results in detail at $b=1$. In Fig. 1. below we present 
a plot of the imaginary part of $\gamma$ as a function of Reynolds number. 
In Fig. 2 below we plot the real part of the same frequency as a
a function of Reynolds number. Notice that the imaginary part changes sign 
(indicating a transition to instability) at a Reynolds number of order 13. 

We have performed the same analysis at several different values of $b$. 
At each value of $b$ the imaginary part of $\gamma$ turns negative at 
a particular value of the Reynolds number. In Fig. 3 below we have plotted 
this critical Reynolds number as a function of $b$ for a range of values of 
$b$. 

\begin{figure}[ht!]
 \begin{center}
\includegraphics[scale=0.9]{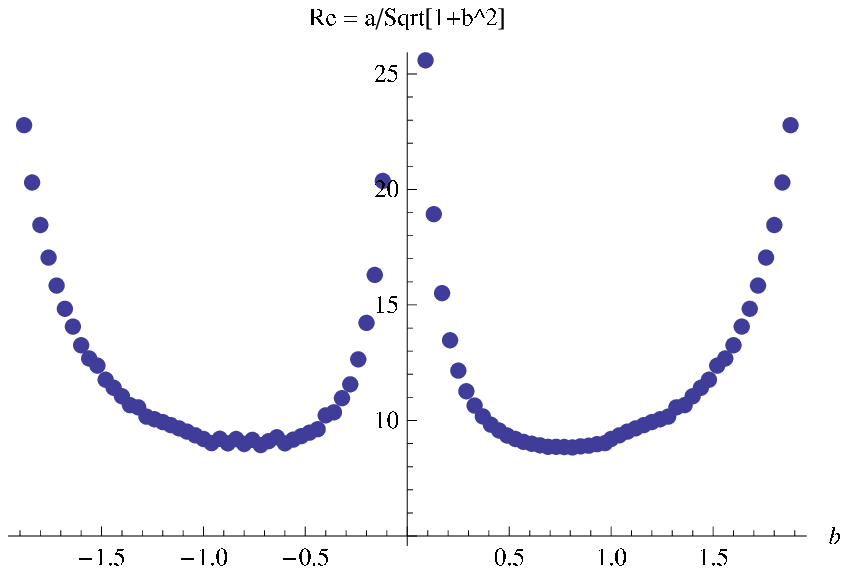}
\end{center}
\caption{The curve that separates the region of stability (below) from 
instability (above) of the eigenvalue at $K=1$,  plotted as a function of 
the Reynolds number on the $y$ axis versus $b$ on the $x$ axis. We have 
used $\kappa=0$, $c=1$ in the calculations that generated this plot.}
\label{rei}
\end{figure}

As our principle aim in this subsection is to establish that the small 
fluctuations about our solutions are unstable at high enough Reynolds numbers
we have not attempted to carefully estimate the errors in our 
numerical calculations. However we think it is unlikely that the errors 
in, for instance, Fig. 3 exceed a few percent. 

As we have described above, at every value of $b$ we start our calculations 
at small Reynolds numbers which are then slowly increased. As a check on 
our numerics, at small  we have compared our numerical results for $\gamma$ 
versus $k$ against the predictions of perturbation theory (see the next 
subsection for details)  at lowest order in $(Re)^2$; we find good agreement.
In fact at $b=1$ and $a=1/100$, the difference between our numerical result 
and the zeroth order answer matched the prediction of perturbation theory 
upto one part in $10^5$; a more accurate result than we had expected.

\subsection{Perturbation theory at small Reynolds numbers}

Let us formally identify the equation \eqref{recrelf} with free index $k$ 
with 
$$\langle k| M |\psi \rangle= \sum_m \langle k | M | m\rangle \langle m| 
\psi \rangle$$
and simultaneously identify 
$$ \langle k| \psi \rangle = f_k$$
With these formal identifications, the set of equations \eqref{recrelf}
(for all $k$) are simply the equation for the operator $M$ to have 
a zero eigenvalue. The corresponding eigenvector is specified by the 
values of $f_k$ on this solution.

Let us write $M = M_0 +  M_1$ where $M_0$ is a diagonal operator in the 
basis described above and  $M_1$ is an off-diagonal operator, proportional 
to $\alpha$.  The elements of the matrix $M_0$ and $M_1$ are given by  
\begin{equation}\label{component}
\begin{split}
\langle f_k|M_0|f_k\rangle &= A(k) \\
              & = \left[ \frac{i}{b}  (\gamma + k)((\kappa+k)^2 + c^2) + 
((\kappa+k)^2+c^2)^2 \right]\delta_{kj}\\
\langle f_k|M_1|f_{k-1}\rangle &= B(k)\\
                    &= i c a \left( -\frac{(\kappa+k -1)^2+ c^2-1}
{1-ib} +1 \right)\delta_{k,(j-1)} \\
\langle f_k|M_1|f_{k+1}\rangle &= C(k)\delta_{k,(j+1)}\\
                         &=  i c a^* \left( -\frac{(\kappa+k +1)^2+ c^2-1}
{1+ib} +1 \right)\delta_{k,(j+1)}
\end{split}
\end{equation}

When $\alpha=0$ the eigenvectors of the operator $M$ are simply $|K\rangle$
with eigenvalue $A(K)$. This eigenvalue vanishes when 
$$ \gamma_0 = -K + i~b\left[(\kappa +K)^2 + c^2\right] .$$
We now wish to study how this eigenvector - and the corresponding solution 
for $\gamma$ that keeps the eigenvalue zero - evolves in perturbation 
theory in $\alpha$. 

To lowest nontrivial order in perturbation theory, the shift in the 
eigenvalue $A(k)$ is given by 
\begin{equation}\label{eicor}
\begin{split}
\tilde A(k) &= A(k,\omega) + \frac{\langle f_k|M_1|f_{k-1}\rangle\langle f_{k-1}|M_1|f_k\rangle}{A(k) - A(k-1)} + \frac{\langle f_k|M_1|f_{k+1}\rangle\langle f_{k+1}|M_1|f_k\rangle}{A(k) - A(k+1)}\\
&=A(k) + \frac{B(k)C(k-1)}{A(k) - A(k-1)} + \frac{C(k)B(k+1)}{A(k) - A(k+1)}
\end{split}
\end{equation}
In order that the eigenvalue vanish at this order we must have 
$$\gamma = -K  + i~b\left[(\kappa +K)^2 + c^2\right] -\frac{i~b}{(K+ \kappa)^2 + c^2}\bigg[ \frac{B(K)C(K-1)}{A(K-1)} + \frac{C(K)B(K+1)}{ A(K+1)}\bigg]$$
where the third term is evaluated at $\gamma = \gamma_0$ and we have 
used the fact that $A(K)=0$ at $\gamma=\gamma_0$. 

\section{A simple flow on a sphere}\label{app:sphere}

It is not difficult to generalize the simple laminar flow presented 
in section 4 above to a shear flow on a 2 sphere. The corresponding dual 
gravitational solution 
to this flow is asymptotic to (slightly perturbed) global AdS space.  

For concreteness we choose $d=3$ and choose the 
base metric of our space to be the two sphere, i.e. 
\begin{equation} \label{ts}
ds^2=-dt^2+d\theta^2 + \sin^2 \theta d \phi^2
\end{equation}

Our velocity and forcing function fields will be vector fields on the 
two sphere, so be briefly pause to recall some necessary definitions. 
Recall that the spherical harmonics, $Y^{l}_m$ form a basis for the 
expansion of an arbitrary scalar field on the sphere. On the other 
hand an arbitrary vector field on the sphere is given by 
linear combinations of $\partial_i Y^l_m$ and $\epsilon_i^j\partial_j Y^l_m$
where the $\epsilon$ symbol includes relevant factors of $\sqrt{g}$. Now, 
in our problem, both $v^i$ and $a^i$ are divergenceless. It follows 
that each of these fields may be expanded in a sum over only the vector 
(rather than also the derivative of scalar) spherical harmonics.

In order to obtain one solution (that can easily be generalized in many ways)
to the Navier-Stokes equations, let 
$$a_i= \alpha \epsilon_i^m \partial_m Y^l_0(\theta, \phi).$$
The velocity configuration  \footnote{We use that 
$$\nabla^m \nabla_m \nabla_i \phi
= \nabla_i \nabla^m \nabla_m  + R_i^m \nabla_m \phi$$.}

$$v_i=  - \frac{\alpha}{1- i \frac{l (l+1)-2}{\omega}} \times 
\epsilon_i^m \partial_m Y^l_0(\theta, \phi).$$
(together with the implied pressure field) yields a steady state, laminar, 
shear flow. The Reynolds number of this flow is given by \eqref{re} 
with $a$ and $b$ now given by $a=\frac{\alpha}{k_o \nu}$ and $b=\frac{\nu (l(l+1)-2}{\omega}.$ As for the solution presented in section 4, 
we expect this flow to be unstable to small 
perturbations at fixed $b$ in the limit of large $a$ (we have not 
performed the linear stability analysis in this case).

\bibliographystyle{JHEP}
\bibliography{scanew}

\end{document}